\documentclass{ws-procs975x65}
\usepackage{amsmath}
\usepackage{amsfonts}
\usepackage{epsfig}
\usepackage{figlabel}

\newcommand{\hs}[1]{\hspace*{#1cm}}
\newcommand{\vs}[1]{\vspace*{#1cm}}

\newcommand{\la}{\langle}
\newcommand{\ran}{\rangle}
\newcommand{\half}{{\textstyle \frac{1}{2}}}

\newcommand{\bbeta}{\mbox{\boldmath $\beta$}}
\newcommand{\bgamma}{\mbox{\boldmath $\gamma$}}

\newcommand{\bh}{\mbox{\boldmath $h$}}

\newcommand{\grad}{\nabla}

\begin{document}

\title{Simultaneity and the Concept of `Particle'}
\author{Carl E. Dolby}
\address{Department of Theoretical Physics, 1 Keble Rd, Oxford OX1 3RH, U.K.\\E-mail: dolby@thphys.ox.ac.uk} 

\maketitle

\abstracts{
The history of the particle concept is briefly reviewed, with
particular emphasis on the `foliation dependence' of many
particle creation models, and the possible connection between our notion
of particle and our notion of simultaneity. It is argued that the
concept of `radar time' (originally introduced by Sir Hermann Bondi in his
work on k-calculus) provides a satisfactory concept of `simultaneity'
for observers in curved spacetimes. This is used to propose an
observer-dependent particle interpretation, applicable to an
arbitrary observer, depending solely on that observers motion and not
on a choice of coordinates or gauge. This definition is illustrated
with application to non-inertial observers and simple cosmologies,
demonstrating its generality and its consistency with known cases.}

\section{Introduction}

	In this conference we have heard illuminating discussions of 
various aspects of the role of time in physics, and the conceptual 
tension that often surrounds it. One well-known tension 
is between the `effectively absolute' role that time plays in quantum 
mechanics, and the role it plays in general relativity, where it is 
just one coordinate in a covariant theory. My contribution to 
these proceeding will discuss the problem of particle creation in gravitational 
backgrounds, and in accelerating reference frames. In so doing I hope to 
shed some light on the aforementioned tension, and also to describe a 
fascinating connection that exists between our concept of simultaneity, 
and our concepts of `particle' and `vacuum'.

	The first prediction of particle creation in gravitational 
backgrounds came in 1939 when Schr\"{o}dinger\cite{Schrod} predicted 
that if the universe is expanding then ``it would mean production of 
matter merely by its expansion''. This prediction was 
readressed in detail by Parker\cite{Park1,Park2,Park3} in the late 60's. However, 
gravitational particle creation first hit the headlines in 1975, with the 
discovery of Hawking 
radiation from black holes \cite{Haw2}. Perhaps an even more intriguing 
discovery was made later that year, by Unruh\cite{Unruh} and 
independently by Davies\cite{Davies}. They showed that an 
observer who accelerates uniformly through flat empty space will also 
observe a thermal bath of particles, at a temperature given by their 
acceleration. This means that a state which is empty according to an 
inertial observer will not be empty according to an accelerating observer, 
and hence it demonstrates that the concept of `vacuum' (and hence of 
`particle') must be observer-dependent. 
	To see how these predictions could arise, consider a globally 
hyperbolic spacetime, and for definiteness, consider massive Dirac 
fermions. Then we have a field operator $\hat{\psi}(x)$ 
satisfying\cite{GMR,Kak}:

\begin{equation} (i \gamma^{\mu} \grad_{\mu} - m) \hat{\psi}(x) = 0 \label{gov} 
\end{equation}

where $\{ \gamma^{\mu} , \gamma^{\nu} \} = 2 g^{\mu \nu}$, 
$\grad_{\mu} \psi \equiv \partial_{\mu}\psi + \Gamma_{\mu}\psi$ 
and $\Gamma_{\mu} = \frac{1}{4} \gamma_{\nu} \grad_{\mu} \gamma^{\nu}$ is 
the spin connection. Since no interactions are present then we can 
expand $\hat{\psi}(x)$ in terms of a complete set of normal modes as:

\begin{equation} \Longrightarrow  \hat{\psi}(x) = \sum_{i}  
\{ u_{i}(x) a_{i} + v_{i}(x) b^{\dagger}_{i} \} \label{eq:psihatt0} \end{equation}

However, these modes are no longer simple plane waves, so it is no 
longer obvious which modes should be put with the $a_{i}$ operators and 
interpreted as particle modes, and which should be put with the 
$b_{i}^{\dagger}$ operators and interpreted as antiparticle modes. Two choices 
must be made. The `in' modes $\{ u_{i,\rm{in}}, v_{i,\rm{in}} \}$, chosen to 
represent particle/antiparticle modes at early times, determine the `in' vacuum 
$|0_{\rm{in}}\ran$ by the requirement:

$$ a_{i,\rm{in}} |0_{\rm{in}}\ran = 0 = b_{i,\rm{in}} |0_{\rm{in}}\ran $$

The `out' modes $\{ u_{i,\rm{out}}, v_{i,\rm{out}} \}$ determine the `out' 
number operator 
$$\hat{N}_{\rm{out}} = \sum_{i} \{ a^{\dagger}_{i,\rm{out}} a_{i,\rm{out}} + b^{\dagger}_{i,\rm{out}} b_{i,\rm{out}} \}$$

By expanding the `out' modes in in terms of the `in' modes we obtain:
\begin{align} a_{i,\rm{out}} & = \sum_j \{ \alpha_{i j} a_{j,\rm{in}} + 
\beta_{i j} b^{\dagger}_{j,\rm{in}} \} \label{eq:new1} \\
 b_{i,\rm{out}} & = 
\sum_j \{ \gamma_{i j}^* a^{\dagger}_{j,\rm{in}} +
\epsilon_{i j}^* b_{j,\rm{in}} \} \label{eq:new2} \end{align}

The number of `out' particles in the `in' vacuum is then given by:

\begin{equation}  \la 0_{\rm{in}} | \hat{N}_{\rm{out}} | 0_{\rm{in}} \ran = Trace( \bbeta \bbeta^{\dagger} + \bgamma \bgamma^{\dagger}) \label{ntot1} \end{equation}
hence describing particle creation. The task of describing particle creation then 
boils down to the question: {\it How do we choose the `in' and `out' modes?}

There are a large variety of methods proposed for this choice (see for 
instance the common texts\cite{BD,Full} and the references therein), 
based on adiabatic expansions, conformal symmetry, killing vectors, the 
diagonalisation of a suitable Hamiltonian, or many other methods. Broadly 
speaking these methods are limited by one of two drawbacks. Either they 
require the spacetime to possess certain desirable symmetries (deSitter, Killing 
vectors, conformal symmetries etc), or they give results which depend on an 
arbitrary foliation of spacetime into `space' and `time'. Meanwhile, although 
a choice of observer often motivates the choice of foliation (such as in 
the Unruh effect), there is no systematic prescription for linking the chosen 
observer to the chosen foliation.

Many of these drawbacks can be avoided by introducing a model particle 
detector\cite{Unruh,BD,Tak,Sr1}. This provides an operational particle 
concept, which directly incorporates the observers motion. However, it can 
not be used to define the particle/antiparticle modes, for a number of 
reasons. Firstly, because a detector only counts particles on its trajectory, 
so could not for instance categorize the emptiness of a state. More 
importantly, it would be circular. Provided a particle detector is anything 
that detects particles, a particle cannot also be ``anything detected by a 
particle detector''. Even if we stick only to `tried and tested' detector 
models\cite{Unruh,DeWitt}, then the question arises ``what were they tested 
against?'' - we must have in mind a concept of particle before fashioning a 
concept of detector. (In the case of fermions there are also technical 
difficulties\cite{Tak,Sr1,Sr2}, meaning that the predictions of current detector 
models are not always proportional to the number of particles present, even 
for inertial detectors in electro-magnetic fields.)

	In this article we offer a resolution to these 
difficulties\cite{mythesis,Dolby,Dolby3} which builds on 
the so-called `Hamiltonian diagonalisation' prescription\cite{GM1,GM2,MMS,GMM}; a method
 criticized in the past\cite{Full2} for its reliance on an arbitrarily chosen
 foliation of spacetime (time coordinate). Our resolution lies in using the
concept of `radar time\footnote{Also known\cite{Pauri} as ``M\"{a}rzke-Wheeler coordinates''.}'
 (originally introduced by Sir Hermann Bondi\cite{Bondi,Bohm,Me3} 
in his work on k-calculus) to uniquely assign a foliation of spacetime to any 
given observer. The result is a particle interpretation which depends {\it only} 
on the motion of the observer, and on the background present, and which 
generalizes Gibbons' definition\cite{Gibb2} to arbitrary observers and 
non-stationary spacetimes. It also facilitates the definition of a number 
density operator, allowing us to calculate not just the total asymptotic 
particle creation, but also to say (with definable precision) where and when 
these particles were `created'.

Given the central role that radar time will play in this particle interpretation, 
the next Section is devoted to describing radar time, while Section 3 describes 
the application of radar time to an arbitrary observer in 1+1 Dimensional Minkowski 
Space. The observer-dependent particle interpretation is defined and discussed 
in Section 4. In Section 5 we return to 1+1 Dimensional Minkowski space, and 
describe the massless Dirac Vacuum as seen by an arbitrarily moving 
observer. Conclusions are presented in Section 6.

\section{Radar Time}

Consider an observer traveling on path $\gamma: x^{\mu} = x^{\mu}(\tau)$ 
with proper time $\tau$, and define:
\begin{align}
\tau^{+}(x) & \equiv \mbox{ (earliest possible) proper time at which a 
null geodesic leaving} \notag \\
& \hs{2} \mbox{ point $x$ could intercept $\gamma$. } \notag \\
\tau^{-}(x) & \equiv \mbox{ (latest possible) proper time at which a null 
geodesic could} \notag \\
& \hs{2} \mbox{ leave $\gamma$, and still reach point $x$. } \notag \\
\tau(x) & \equiv \half (\tau^{+}(x) + \tau^{-}(x)) \hs{1} = 
\mbox{ `radar time'.} \notag \\
\rho(x) & \equiv \half (\tau^{+}(x) - \tau^{-}(x)) \hs{1} = 
\mbox{ `radar distance'.} \notag \\
\Sigma_{\tau_0} & \equiv \{x: \tau(x) = \tau_0 \} = \mbox{ observer's `hypersurface 
of simultaneity at time $\tau_0$'. } \notag \end{align}

\begin{figure}[h]
\vspace{-.3cm}
\center{\epsfig{figure=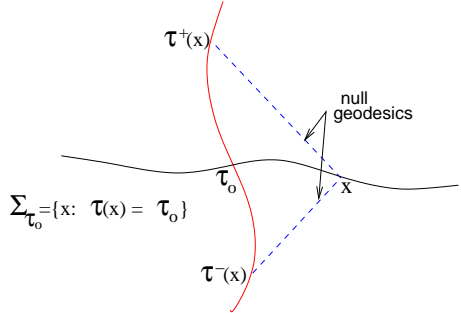 , width=6cm}}
\caption{{\footnotesize Schematic of the definition of `radar time' $\tau(x)$.}}
\vspace{-.3cm}
\end{figure}

This is a simple generalization of the definition made popular by Bondi in his
 work on special relativity and {\it k}-calculus\cite{Bondi,Bohm,Dinverno}. It 
can be applied to any observer in any spacetime. We can also define the 
`time-translation' vector field: 
\begin{equation} k_{\mu}(x) \equiv \frac{\frac{\partial \tau}{\partial 
x^{\mu}}}{g^{\sigma \nu} \frac{\partial \tau}{\partial x^{\sigma}} 
\frac{\partial \tau}{\partial x^{\nu}}} \label{eq:time} \end{equation}
	This represents the perpendicular distance between neighboring 
hypersurfaces of simultaneity, since it is normal to these hypersurfaces and 
it satisfies $k^{\mu}(x)\frac{\partial \tau}{\partial x^{\mu}} = 1$. Radar time is 
independent of the choice of coordinates, and is single valued in the 
observers {\it causal 
envelope} (the set of all spacetime points with which the observer can both 
send and receive signals). An affine reparametrisation of the observers 
worldline leads only to a relabeling of the same foliation, such that the 
radar time always agrees with proper time on the 
observer's path. It is invariant under `time-reversal' - that is, 
under reversal of the sign of the observer's proper time. 

	We now illustrate these properties with a simple class of examples; observers 
in 1+1 Dimensional Minkowski space. Some simple cosmological examples are presented 
elsewhere\cite{Dolby3}.

\section{Arbitrary Observer in 1 + 1 Dimensions}

Let the observers worldline be described by 
$$x^{\pm} \equiv t \pm x = x^{\pm}_{\lambda}(\tau_{\lambda}) = \int^{\tau_{\lambda}} e^{\pm \alpha(\tau)} {\rm d} \tau$$ 

where $\tau_{\lambda}$ is the observers proper time, and 
$\alpha(\tau_{\lambda})$ is the observers `rapidity' at time $\tau_{\lambda}$. 
$e^{\alpha(\tau_{\lambda})}$ is the obvious time-dependent generalization 
of the `k' of Bondi's k-calculus\cite{Bondi,Bohm,Dinverno}.  The observers 
acceleration is $a(\tau_{\lambda}) = \frac{d \alpha}{d \tau_{\lambda}}$. The 
observers worldline is completely specified by the choice of origin 
(i.e. $x^{\mu}(0)$) and the rapidity function $\alpha(\tau_{\lambda})$, or 
by the choice of origin, the initial velocity, and the function 
$a(\tau_\lambda)$.

	It is straightforward to show that the coordinates $\tau^{\pm} = \tau \pm \rho$ are given by:
$$ x^{\pm} = x^{\pm}_{\lambda}(\tau^{\pm}) $$
while the metric in these coordinates\footnote{For convenience we have 
reversed the role of $\tau^+$ and $\tau^-$ to the 
observers left, so that $\rho$ plays the role of a spatial (rather than radial) 
coordinate, being negative to the observers left - the radar time is unchanged 
by this.} is:
$${\rm d} s^2 = e^{(\alpha(\tau^+) - \alpha(\tau^-))} ({\rm d} \tau^2 - {\rm d} \rho^2) $$ 

 We see that the radar coordinates are obtained from the Minkowski 
coordinates simply by rescaling along the null axes. The 
`time-translation vector field' \cite{Dolby3,Dolby} is simply $k^{\mu} 
\frac{\partial }{\partial x^{\mu} } =  \frac{\partial }{\partial \tau}$, 
while the hypersurfaces $\Sigma_{\tau}$ are hypersurfaces of 
constant $\tau$.

As a useful consistency check, consider an inertial observer with a 
velocity $v $ relative to our original frame. Then $\alpha$ is constant, 
and $x^{\pm}_{\lambda}(\tau_{\lambda}) = e^{\pm \alpha} \tau_{\lambda}$. The 
coordinates $\tau^{\pm}$ are hence given by $\tau^{\pm} = e^{\mp \alpha} x^{\pm}$ so:
$$\tau = \half(e^{- \alpha} x^+ + e^{\alpha} x^-) = \frac{t - v x}{\sqrt{1 - v^2}}, 
\hs{1} \rho = \half(e^{-\alpha} x^+ - e^{\alpha} x^-) = \frac{x - v t}{\sqrt{1 - v^2}}$$
   The radar coordinates of an inertial observer are just the coordinates of 
their rest frame, as expected.

\subsection{Constant Acceleration}

The simplest nontrivial case is constant acceleration. In this case 
$\alpha(\tau) = a \tau$, and we have 
$x^{\pm}_{\lambda}(\tau_{\lambda}) = \pm a^{-1} e^{\pm a \tau_{\lambda}}$ which gives:
\begin{align} \tau & = \frac{1}{2 a} \log\left(\frac{x+t}{x-t}\right) \hs{2} \rho = \frac{1}{2 a} \log\left(a^2(x^2 - t^2)\right) \\
 {\rm d} s^2 & = e^{2 a \rho} ({\rm d} \tau^2 - {\rm d} \rho^2) \end{align}
These are Rindler coordinates, which cover only region U of Figure 2, as expected.
The hypersurfaces of constant $\tau$ are given by $t_{\tau_0}(x) = x \tanh(a \tau_0)$. 

\begin{figure}[h]
\vspace{-.3cm}
\center{\epsfig{figure=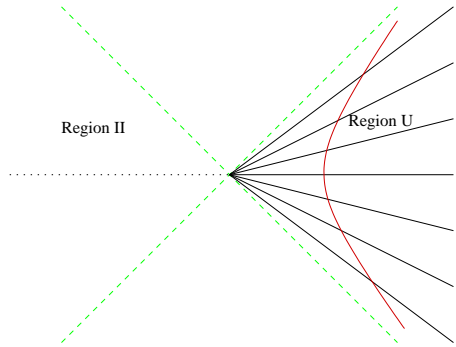 , width=6cm}}
\caption{{\footnotesize Hypersurfaces of simultaneity of a uniformly accelerating observer.}}
\vspace{-.3cm}
\end{figure}

\vs{-.5}

\subsection{Gradual Turnaround Cases}

Consider now an observer (Barbara say) who accelerates uniformly for 
$|\tau_{\lambda}| < \tau_c$, but is otherwise inertial. Then $\alpha(\tau_{\lambda}) 
= a \tau_{\lambda}$ for $|\tau_{\lambda}| < \tau_c$ and $= \pm a \tau_c$ for 
$\tau_{\lambda} > \tau_c$ or $< - \tau_c$ respectively.

\begin{figure}[htb]
\figstep
\begin{minipage}[b]{6.8cm}
\epsfig{figure=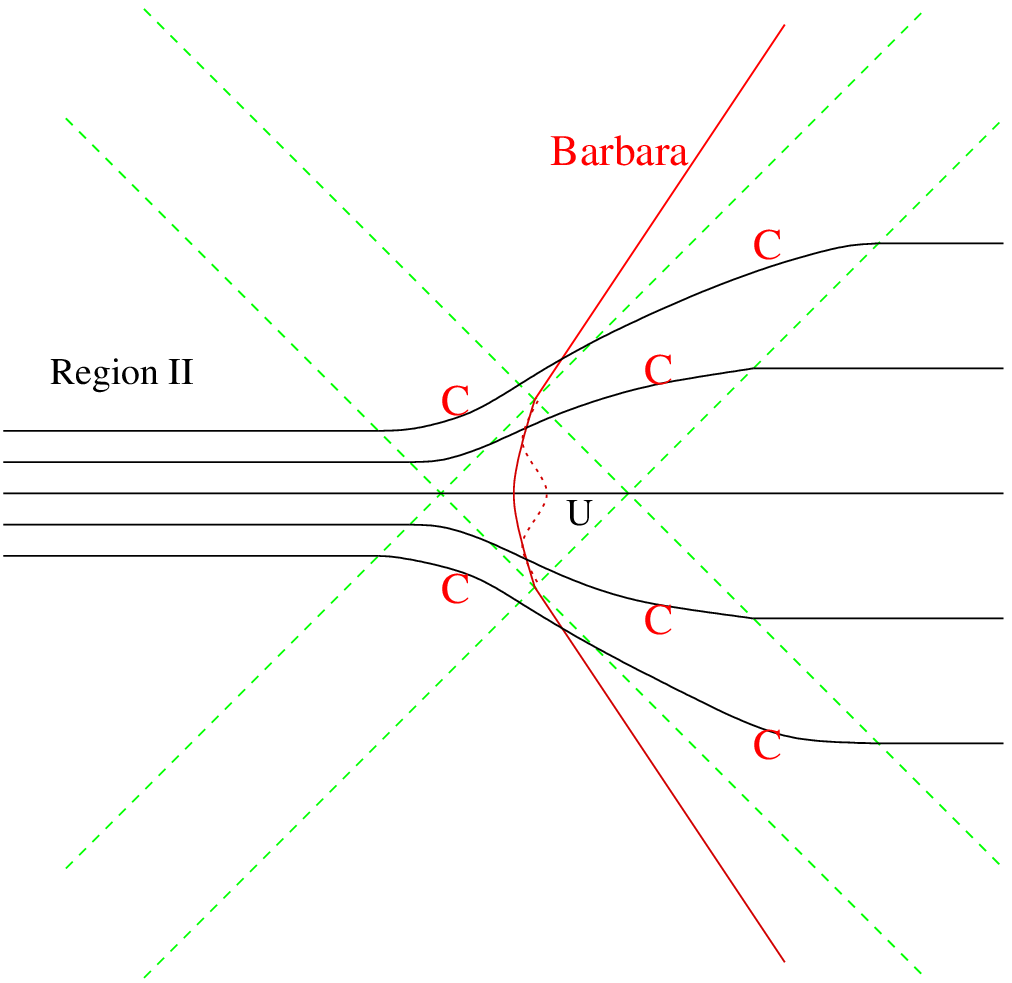, width=6.8cm}

{\footnotesize {\bf Figure \ref{fig1}(A).} Barbara's hypersurfaces of constant $\tau$.} \vs{.2} 
\end{minipage}\hs{1}\begin{minipage}[b]{4.8cm}
\epsfig{figure=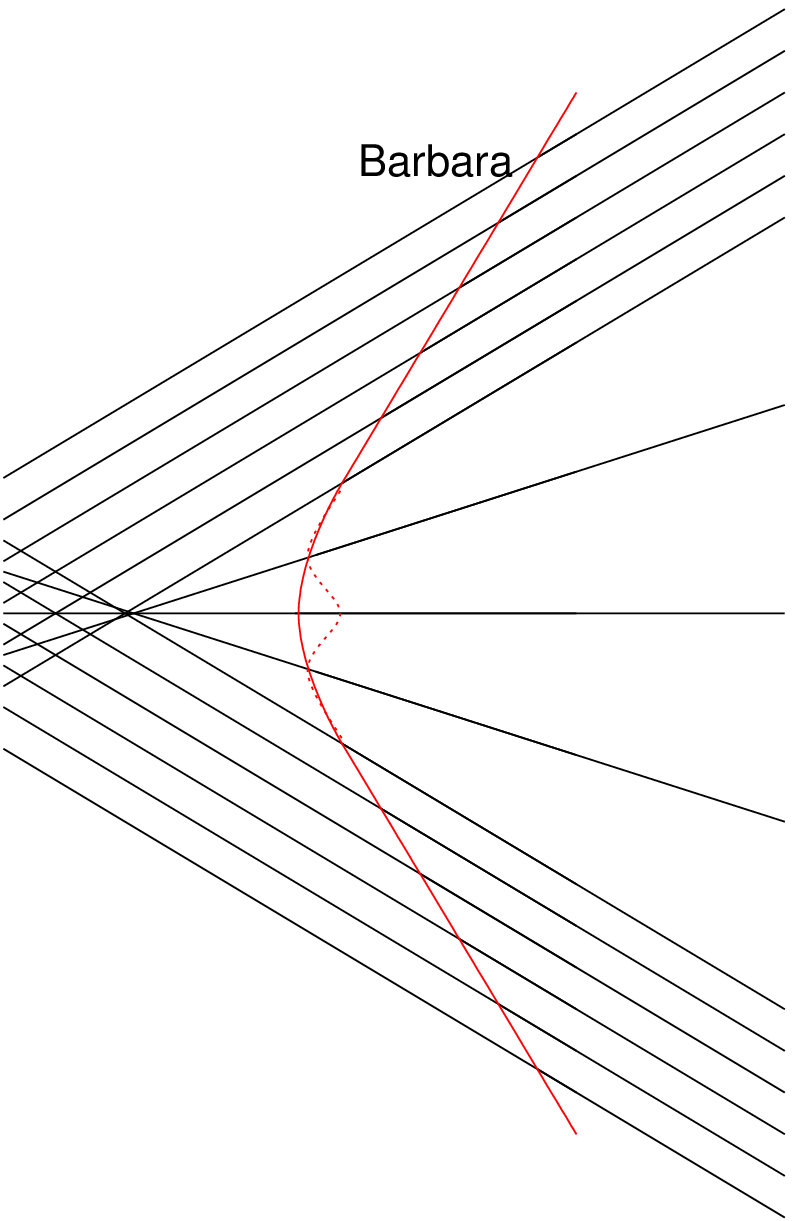, width=4.8cm}

{\footnotesize {\bf Figure \ref{fig1}(B).} Barbara's instantaneous rest frames.}
\end{minipage}
\vs{-.6}
\figlabel{fig1}
\end{figure}

The hypersurfaces of simultaneity for this observer are shown\footnote{This 
is also described elsewhere\cite{Me3}, in the context of the well-known relativistic 
twin ``paradox''.} in Figure 3 (A). For comparison we have included the 
standard `instantaneous rest frame' in Figure 3 (B). The instantaneous rest 
frame suffers from being triple valued to the observers left. It is also 
sensitively dependent on the small-scale details of Barbara's trajectory. Consider 
for instance a small deviation of Barbara's trajectory, like the small dotted line 
in the turnaround point of Figures 3 (A) and 3 (B). In figure 3 (B) this has 
serious effects - Barbara now assigns five times to events far to her left, and 
three to events far to her right! In 3 (A) however, this change causes only a 
small change in the times assigned to events in the vicinity of the points marked $C$.

A similar example is that of an observer with trajectory given by:

$$x^{\pm}_{\lambda}(\tau_{\lambda}) = \int_{0}^{\tau_{\lambda}} e^{\pm a \tau_c \tanh(\tau/\tau_c)} {\rm d} \tau $$

This observer has acceleration $a(\tau) = a \cosh^{-2}(\tau/\tau_c)$, so is uniformly accelerating for $|\tau| << \tau_c$ and inertial for $|\tau| >> \tau_c$. We will return to this example shortly.

\section{An Observer-Dependent Particle Interpretation}

Consider again the field operator:
\begin{equation} \hat{\psi}(x) = \sum_i \{ u_{i,\rm{in}}(x) a_{i,\rm{in}} 
+ v_{i,\rm{in}}(x) b_{i,\rm{in}}^{\dagger} \} \label{field1} \end{equation}
and the state $|\rm{in}\ran$ defined by $a_{i,\rm{in}} |\rm{in}\ran = 0 = 
b_{i,\rm{in}} |\rm{in}\ran$. We will consider the time-dependent particle 
content of this state, as measured by an observer O. We mentioned in the 
introduction that this definition stems from the diagonalisation of a 
suitable Hamiltonian. The Hamiltonian in question is:
\begin{align} \hat{H}(\tau) & = \int_{\Sigma_{\tau}} \sqrt{-g} \ k^{\mu} T_{\mu \nu}(\hat{\psi},\hat{\psi}) {\rm d} \Sigma^{\nu} \\
\mbox{where } T_{\mu \nu}(\psi,\phi) & = \half i [\bar{\psi} \gamma_{(\mu} \grad_{\nu)} \phi - (\grad_{(\mu} \bar{\psi}) \gamma_{\nu)} \phi]  \end{align}
	$T_{\mu \nu}(\hat{\psi},\hat{\psi})$ is the (unregularised) 
stress-energy tensor for Dirac fermions\cite{BD}. Diagonalising this 
Hamiltonian\cite{GMM} entails expanding $\hat{\psi}$ as:
\begin{align}
\hat{\psi}(x) & = \sum_i u_{i,\tau_0}(x) a_{i,\tau_0} + \sum_i v_{i,\tau_0}(x) b^{\dagger}_{i,\tau_0} + \sum_i w_{i,\tau_0}(x) c_{i,\tau_0} \label{field2}\\
 & = \hat{\psi}^{(+)}_{\tau_0}(x) + \hat{\psi}^{(-)}_{\tau_0}(x) + \hat{\psi}^{(0)}_{\tau_0}(x) \end{align}
	and choosing these modes such that the Hamiltonian becomes:
\begin{equation} \hat{H}(\tau_0) = \sum_{i j} h^{(+)}_{i j}(\tau_0) 
a^{\dagger}_{i,\tau_0} a_{j,\tau_0} - \sum_{i j} h^{(-)}_{i j}(\tau_0) 
b_{i,\tau_0} b^{\dagger}_{j,\tau_0} \label{hdiag} \end{equation}
where the matrices $\bh^{(\pm)}$ are positive definite. To consider the 
content of this requirement, it is convenient to define the `1st quantized 
Hamiltonian' $\hat{H}_1(\tau)$ (on the space of finite-norm solutions of the 
Dirac equation) by:
$$ \la \psi | \hat{H}_1(\tau) |\phi\ran = \int_{\Sigma_{\tau}} \sqrt{-g} \ k^{\mu} T_{\mu \nu}(\psi,\phi) {\rm d} \Sigma^{\nu} $$
	Then equation (\ref{hdiag}) requires that $\{ u_{i,\tau_0}(x) \}$ span 
the positive spectrum of $\hat{H}_1(\tau_0)$, $\{ v_{i,\tau_0}(x) \}$ span the 
negative spectrum of $\hat{H}_1(\tau_0)$, and $\{ w_{i,\tau_0}(x) \}$ span the 
null space of $\hat{H}_1(\tau_0)$. The $w_{i,\tau_0}(x)$ will generally 
be states of compact support outside the causal envelope of the 
observer\footnote{However, even for inertial observers in electromagnetic 
backgrounds, there exist topologically non-trivial backgrounds for which zero 
energy eigenstates exist, leading to the existence of fractional charge\cite{Jackiw}. Although such situations 
are straightforward to describe within the present approach, we will not 
discuss them further here.}.
	Having defined $\hat{\psi}^{(\pm)}(x)$ and $\hat{\psi}^{(0)}(x)$ by 
this requirement, we can now define the particle number operator on $\Sigma_{\tau_0}$, $\hat{N}^+_{\tau_0}$ by:

\begin{align} \hat{N}^+_{\tau_0} = \int_{\Sigma_{\tau_0}} & \sqrt{-g} \  
\bar{\hat{\psi}}^{(+)}_{\tau_0} \gamma^{\mu} \hat{\psi}^{(+)}_{\tau_0} 
{\rm d} \Sigma_{\mu} \hs{.5} = \int_{\Sigma_{\tau_0}} \sqrt{-g} \ \hat{J}^{\mu (+)} {\rm d} \Sigma_{\mu} \\
\mbox{ where }  & \hat{J}^{\mu (+)} = \bar{\hat{\psi}}^{(+)}_{\tau(x)} \gamma^{\mu} \hat{\psi}^{(+)}_{\tau(x)} \end{align}

For any state and any chosen observer, the field $\la J^{\mu (+)}\ran$ is a covariant 
vector field, which can be interpreted as describing the `flow of 
particles' as seen by this observer. $\hat{J}^{\mu (+)} {\rm d} \Sigma_{\mu}$ 
represents the number of particles in ${\rm d} \Sigma_{\mu}$. Similarly, the 
antiparticle number operator is given by:

\begin{align} \hat{N}^-_{\tau_0} = - : \int_{\Sigma_{\tau_0}} & \sqrt{-g} \ 
\bar{\hat{\psi}}^{(-)}_{\tau_0} \gamma^{\mu} \hat{\psi}^{(-)}_{\tau_0} 
{\rm d} \Sigma_{\mu} : \hs{.5} = \int_{\Sigma_{\tau_0}} \sqrt{-g} \ \hat{J}^{\mu (-)} {\rm d} \Sigma_{\mu} \\
\mbox{ where }  & \hat{J}^{\mu (-)} = - :\bar{\hat{\psi}}^{(-)}_{\tau(x)} \gamma^{\mu} \hat{\psi}^{(-)}_{\tau(x)}: \end{align}

 The normal-ordering is with respect to the observers particle interpretation 
at the time of measurement (i.e. the $b_{i,\tau}$). These operators allow the 
observer to calculate the total number of particles/antiparticles on 
$\Sigma_{\tau}$ for all $\tau$, and to determine how this particle content 
is distributed throughout $\Sigma_{\tau}$. Although the total number operator 
$\hat{N}_{\tau} = \hat{N}^+_{\tau} + \hat{N}^-_{\tau}$ is necessarily non-local 
(no local operator could possibly be consistent with the Unruh effect) it 
will generally be effectively local\cite{mythesis,Dolby} on scales larger than the 
Compton length $\lambda_c = \frac{h}{m c}$ of the particle concerned. Equating expressions (\ref{field1}) and (\ref{field2}) for 
$\hat{\psi}(x)$ gives:
\begin{align}
a_{i,\tau_0} & = \sum_j \{ \la u_{i,\tau_0} |u_{j,\rm{in}}\ran a_{j,\rm{in}} +  \la u_{i,\tau_0} |v_{j,\rm{in}}\ran b^{\dagger}_{j,\rm{in}} \} \\
b_{i,\tau_0} & = \sum_j \{ \la v_{i,\tau_0} |u_{j,\rm{in}}\ran^* a_{j,\rm{in}} +  \la v_{i,\tau_0} |v_{j,\rm{in}}\ran^* b^{\dagger}_{j,\rm{in}} \} \end{align}
	which allows us to deduce for instance:
\begin{align} \la \rm{in} | \hat{N}^+_{\tau_0} | \rm{in} \ran & = \rm{Trace}(\bbeta \bbeta^{\dagger}) \\
\la \rm{in} | \hat{N}^-_{\tau_0} | \rm{in} \ran & = \rm{Trace}(\bgamma \bgamma^{\dagger}) \\
\mbox{ where } \beta_{i j}(\tau_0)   \equiv \la u_{i,\tau_0} |v_{j,\rm{in}}\ran \hs{1} & \mbox{and } \hs{.5} \gamma_{i j}(\tau_0)  \equiv \la v_{i,\tau_0} |u_{j,\rm{in}}\ran 
\label{betgam} \end{align}
	as in equation (\ref{ntot1}). Note that, in the presence of 
horizons, the observer $O$ cannot define a unique `vacuum state' at any 
time $\tau_0$. All he can say is that ``a state $|0,\tau_0\ran$ is vacuum 
throughout $\Sigma_{\tau_0}$'' if:
$$ a_{i,\tau_0} | 0,\tau_0\ran = 0 = b_{i,\tau_0} | 0,\tau_0\ran$$ 
for all $i$. This condition is not unique, since we have said nothing 
about $c_{i,\tau_0} | 0,\tau_0\ran$. This is a natural limitation 
however; since $O$ cannot communicate with points outside his causal 
envelope, we can't expect him to be able to determine particle content 
in such regions.

	Although we have specified the `out' modes $\{ u_{i,\tau}, v_{i,\tau} \}$ 
for all possible `out times', we have not yet discussed the choice of `in' modes 
$\{u_{i,\rm{in}}, v_{i,\rm{in}} \}$. This choice is largely a question of 
convenience, and depends on what state we wish to consider the properties 
of. In the absence of particle horizons (when $\hat{\psi}^{(0)} = 0$) we 
may wish the `in' state to be our observers `in-vacuum' $|0,\tau_{\rm{in}}\ran$ 
prepared at some `in' time $\tau_{\rm{in}}$. Alternatively, we may wish that 
the state $|\rm{in}\ran$ be prepared by someone other than the observer. This 
will be the case shortly, where the content of the inertial vacuum will be 
studied by an accelerating observer. Or we may wish (as is common in cosmological 
applications) to consider a state $|\rm{in}\ran$ which is never considered 
`empty' by any observer, but is instead justified by symmetry considerations\cite{BD}. 

\section{The Massless Dirac Vacuum in 1+1 Dimensions}

	As a concrete example of these definitions, consider now the 
massless Dirac vacuum $|0_M\ran$ of flat 1+1 Dimensional Minkowski space, 
as measured by an arbitrarily moving observer (more detail is presented elsewhere\cite{Mark}). Then the `in' modes are the 
plane wave states, which can be written in the massless case as:
\begin{equation} 
u_{p,\pm,\rm{in}}(x) = e^{-i p x^{\mp}}\phi_{\pm}, \hs{1} v_{p,\pm,\rm{in}}(x) 
= e^{i p x^{\mp}}\phi_{\pm} \hs{1} \mbox{ for } p>0 \label{plane}\end{equation}
where the subscript $\pm$ denotes forward/backward moving modes, and the basis 
spinors $\phi_{\pm}$ satisfy $\bar{\gamma}_1 \bar{\gamma}_0 \phi_{\pm} = \pm 
\phi_{\pm}$ where $\bar{\gamma}_{\mu}$ are the flat space Dirac matrices in 1+1 
Dimensions. It can be shown\cite{Mark} that the modes:
\begin{equation} u_{\omega,\pm,O} = e^{\pm \half \alpha(\tau^{\mp})} e^{-i \omega 
\tau^{\mp}} \phi_{\pm} \hs{1} v_{\omega,\pm,O} = e^{\pm \half \alpha(\tau^{\mp})} e^{i \omega \tau^{\mp}} \phi_{\pm} \hs{1} \mbox{ for } \omega > 0 \label{modes}\end{equation}
 diagonalise $\hat{H}(\tau)$ for all $\tau$. Substituting (\ref{plane}) 
and (\ref{modes}) into (\ref{betgam}) and calculating the integral over $p>0$ 
that is implicit in the Trace, gives:

\begin{align} (\beta \beta^{\dagger})_{\omega \omega',\pm} = 2 \int_{-\infty}^{\infty} & {\rm d} \tau_a e^{-i \omega_d \tau_a} \int_{0}^{\infty} {\rm d} \tau_d \sin(\omega_a \tau_d) g_{\pm}(\tau_a,\tau_d) \hs{.5} = (\gamma \gamma^{\dagger})_{\omega \omega',\pm}^* \\
\mbox{ where } g_{\pm}(\tau_a,\tau_d) & = \frac{1}{\tau_d} - 
\frac{\exp\left(\frac{\mp 1}{2} (\alpha(\tau_a + \tau_d/2) + \alpha(\tau_a - \tau_d/2))\right)}{\int_{- \tau_d/2}^{\tau_d/2} \exp(\mp\alpha(\tau_a + \tau)) {\rm d} \tau } \label{gfunction}\\
\omega_a & = \half(\omega + \omega') \hs{.5} \mbox{ and } \hs{.5} \omega_d = \omega' - \omega \end{align}

From these we can deduce that\cite{Mark} the distribution $n^+_{F}(x)$ of forward 
moving particles exactly matches the distribution of forward moving antiparticles, 
and is given by:

\begin{align} n_{F}(\tau^-) & = \int_{-\infty}^{\infty} \frac{{\rm d} \tau}{2 \pi 
\tau} g_+(\tau^- + \tau,\tau) \hs{.5} = \int_{0}^{\infty} \frac{{\rm d} \omega_a}{2 \pi} n_{F,\omega}(\tau^-) \label{spatialdist} \\
\mbox{where } n_{F,\omega}(\tau^-) & = \int_{-\infty}^{\infty} {\rm d} \tau_a \ 
\frac{\sin[\omega (\tau^- - \tau_a)]}{\pi (\tau^- - \tau_a)} 
\int_{0}^{\infty} {\rm d} \tau_d \ \sin(\omega \tau_d) g_{+}(\tau_a,\tau_d) 
\label{freqdist}\end{align}

This is a function only of $\tau^- = \tau-\rho$, as would be expected for 
forward-moving massless particles. It is defined such that $n_{F}(\tau^-) {\rm d} \rho$ 
gives the number of particles within ${\rm d} \rho$ of the point $(\tau,\rho)$. The 
function $n_{F,\omega}(\tau^-)$ can 
be interpreted as the frequency distribution of forward moving 
particles/antiparticles at the point $\tau^-$. Equation (\ref{freqdist}), 
together with (\ref{gfunction}),expresses this distribution anywhere in the 
spacetime, directly in terms of the observers rapidity. Similarly, the spatial 
distribution of backward-moving particles matches 
that of backward moving antiparticles, and can be defined by $n_{B}(\tau^+) = 
\int_0^{\infty} \frac{{\rm d} \omega}{2 \pi} n_{B,\omega}(\tau^+)$ where the 
expressions for $n_{B,\omega}(\tau^+)$ and $n_{B}(\tau^+)$ are as above, 
but with $g_+$ replaced with $g_-$ and $\tau^-$ replaced with $\tau^+$. Notice
 that if the observers worldline is time-symmetric about $\tau_{\lambda}=0$ 
then $\alpha(-\tau_\lambda) = - \alpha(\tau_{\lambda})$ which gives 
$n_{F,\omega}(\tau') = n_{B,\omega}(-\tau')$ for all $\tau'$. On the 
hypersurface $\tau=0$ for instance, where $\tau^{\pm} = \pm \rho$ this 
implies that the distribution of forward moving particles exactly matches 
that of backward moving particles. On the observers worldline on the other 
hand, the distribution of forward moving particles is the time-reverse of 
that for backward moving particles.

	As a consistency check, consider again an inertial observer. Then 
$\alpha$ is constant, so $g_{\pm}(\tau_a,\tau_d) = 0$, and the particle content 
is everywhere zero, as expected. We now consider other examples.

\subsection{Constant Acceleration}

	For a uniformly accelerating observer we have:
$$ g_+(\tau_a,\tau_d) = g_-(\tau_a,\tau_d) = \frac{1}{\tau_d} - \frac{a}{2 \sinh\left(\frac{a \tau_d}{2}\right)} $$
which is independent of $\tau_a$. Hence the forward and backward moving 
particles are each distributed uniformly in $\rho$ for all $\tau$, and the 
frequency distribution is everywhere given by:

\begin{align} n_{F,\omega} & = n_{B,\omega} = 2 \int_0^{\infty} d \tau_d \ \sin(\omega \tau_d) \left( \frac{1}{\tau_d} - \frac{a}{2 \sinh\left(\frac{a \tau_d}{2}\right)}\right) \\
& = \frac{2 \pi}{1 + e^{\frac{2 \pi \omega}{a}}} \label{thermal} \end{align}
which is a thermal spectrum at temperature $T = \frac{a}{2 \pi k_B}$, as expected.

\begin{figure}[htb]
\figstep
\begin{minipage}[b]{6cm}
\epsfig{figure=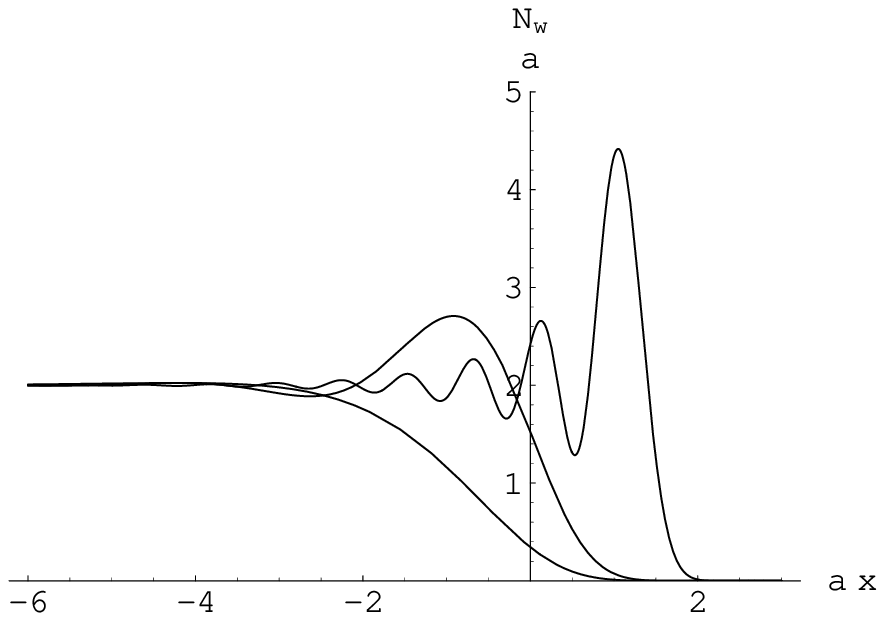, width=6.5cm}

{\footnotesize {\bf Figure \ref{fig2}(A).} $n_{\omega}(\rho)/a$ as a function of $a \rho$, for $m = a$ and $\omega = a/4$ (lowest curve), $a$, and $4 a$ (most oscillatory curve).}
\end{minipage}\hs{.5}\begin{minipage}[b]{6cm}
\epsfig{figure=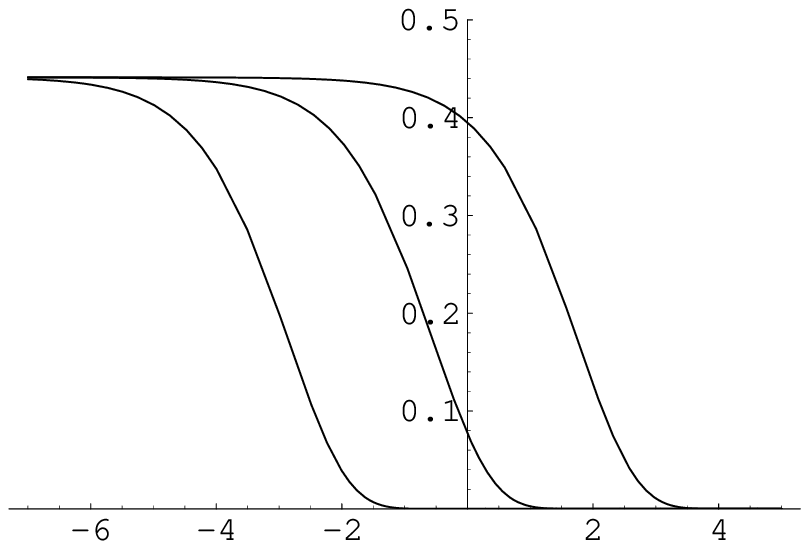, width=7cm}

{\footnotesize {\bf Figure \ref{fig2}(B).} $n(\rho)/a$ as a function of $a \rho$, for $m = a/10$ (right curve), $a$ and $10 a$ (left curve).}
\vs{.07}
\end{minipage}
\figlabel{fig2}
\end{figure}

	For comparison, briefly consider the case of massive fermions in 
1+1 Dimensions (which no longer decompose into forward/backward moving 
modes). In this case the spatially averaged frequency distribution\cite{GMR,Tak,Soffel,Dolby3} is as in (\ref{thermal}), but the massive particles are 
no longer distributed uniformly in $\rho$ (although the distribution is 
completely independent of $\tau$). Nor is the spatial distribution independent
 of $\omega$. Figure 4 shows the spatial distribution of Rindler particles in 
this case\cite{Dolby3}. Figure 4 (A) shows 
$n_{\omega}(\rho)/a$ as a function of $a \rho$ for $m = a$ and $\omega = a/4$
 (lowest curve), $a$, and $4 a$ (most oscillatory curve), while Figure 4 (B)
 shows $n(\rho)/a$ as a function of $a \rho$, for $m = a/10$ (right curve), 
$a$ and $10 a$ (left curve). These can be understood by considering that 
these particles see an `effective mass gap' of $2 m e^{a \rho}$. Each 
frequency penetrates to a value of $\rho$ for which $m e^{a \rho} \sim 
\omega$. Changing the ratio $m/a$ is equivalent to a translation in 
$\rho$. We see that in general the particle number density is uniform to 
the observer's left and negligible to the observer's right, with the 
transition happening at $\rho \sim \frac{1}{a} \log\left(\frac{a}{m}\right)$. 
As $m \rightarrow 0$ this transition point goes to $\infty$, reproducing the 
spatial uniformity 
of the massless limit. However, for non-zero $m$ and realistic accelerations, 
the particle density is small even at low $\rho$ (where it is $\propto a$), while 
the transition to a negligible density occurs far to the 
observer's left.

\subsection{Gradual Turnaround Observer}

Returning to the massless case, consider now the observer with acceleration 
$$a(\tau_{\lambda}) =  a \cosh^{-2}\left(\frac{\tau_{\lambda}}{\tau_c}\right)$$
Their rapidity is $\alpha(\tau_{\lambda}) = a \tau_c \tanh(\tau/\tau_c)$. They 
are accelerating uniformly for $|\tau| << \tau_c$, but are inertial at 
asymptotically early and late times (with velocity $\pm \tanh(a \tau_c)$). There 
are no particle horizons in this case; the observers causal envelope covers 
the whole spacetime. 
 By substituting the rapidity into equation (\ref{gfunction}) we immediately 
obtain the spatial distribution of forward or backward moving particles. At 
time $\tau=0$ these distributions are equal. They are shown in Figure 5, as 
a function of $a \rho$ for $a \tau_c = 1$ (bottom curve), $3, 10, 30, 100$ and $\infty$ 
(top line). As $\tau_c$ increases the particle density increases, and approaches 
the spatial uniformity of the $\tau_c \rightarrow \infty$ limit. 

\begin{figure}
\center{\epsfig{figure=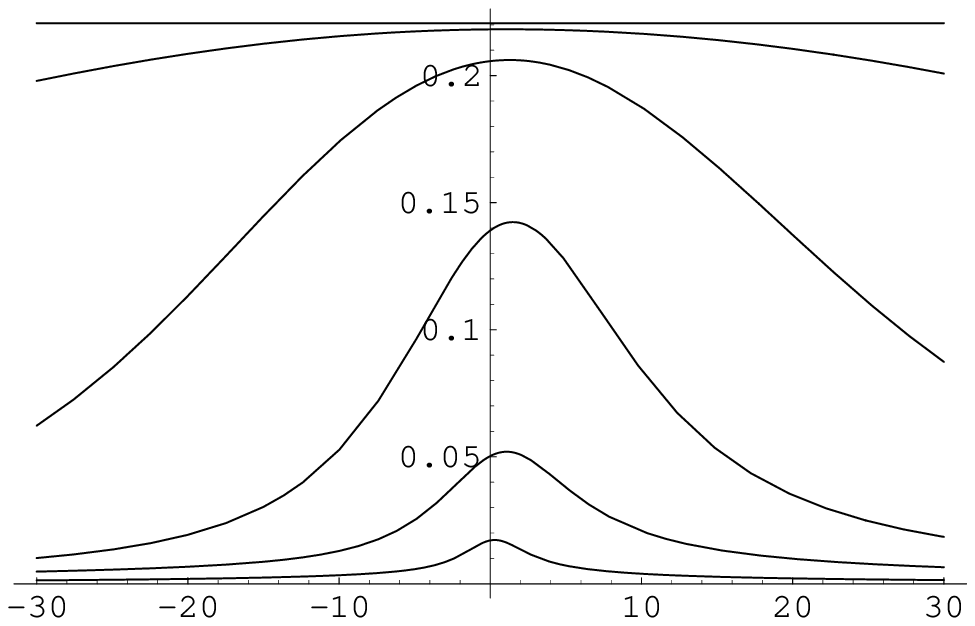, width=8cm}}
\caption{{\footnotesize $n_{F,B}(\rho)/a$ as a function of $a \rho$ for $\tau=0$ and $a \tau_c = 1$ (bottom curve), $3, 10, 30, 100$ and $\infty$ (top curve).}}
\end{figure}

In Figure 6 we have shown the frequency distribution $n_{F,\omega}(\rho)$ of 
forward-moving particles, as a function of $\omega/a$ for $\tau=0$. In Figure 
6 (A) we have $\rho=0$ and $a \tau_c = 3, 10, 30$ and $\infty$. We can clearly 
see that the distribution approaches thermal as $\tau_c$ is increased. In Figure 
6 (B) $a \tau_c = 10$ and $a \rho = \pm 10$. We have also included a plot 
corresponding to a thermal spectrum appropriate to a constant acceleration of 
$a \cosh^{-2}(1)$. The difference between the actual spectrum and the thermal 
spectrum is more significant here. Since $n_{F,B}$ depend only on $\tau^{\pm}$ 
respectively, then Figure 6 (B) also represents the distribution of 
forward/backward moving particles on the observers worldline, at 
$\tau_{\lambda} = 10/a$. We see that the observer sees a different 
number of forward moving particles than backward moving particles. The 
forward/backward moving distributions are the reverse at $\tau=-10/a$.

\begin{figure}[htb]
\figstep
\begin{minipage}[b]{6cm}
\epsfig{figure=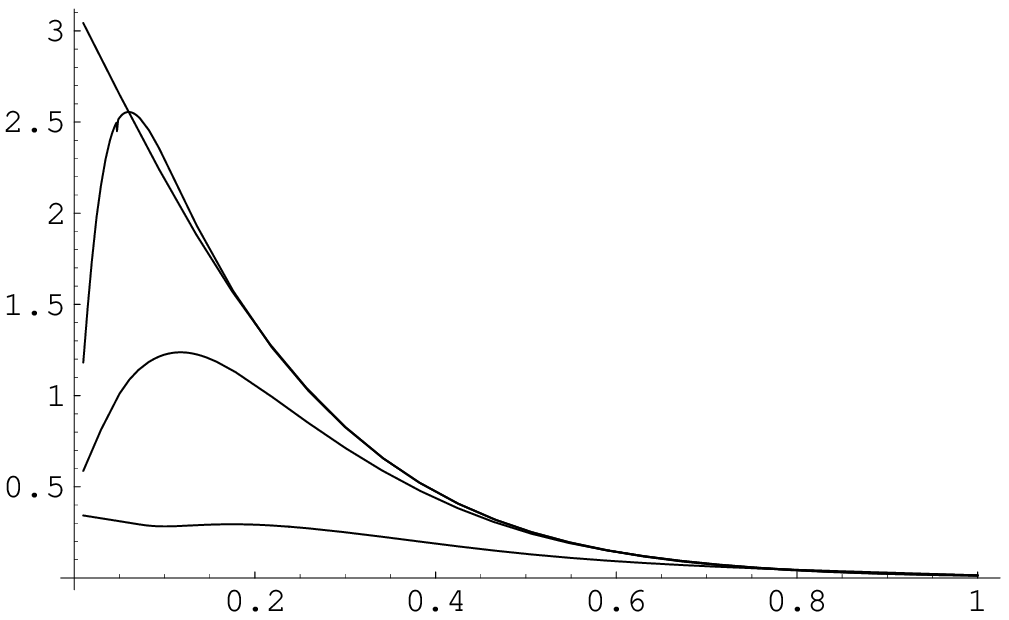, width=6cm}

{\footnotesize {\bf Figure \ref{fig3}(A).} $n_{F,\omega}(\rho)$ as a function of $\omega/a$ for $\rho = 0 = \tau$, and $a \tau_c = 3, 10$ and $30$.}
\end{minipage}\hs{.5}\begin{minipage}[b]{6cm}
\epsfig{figure=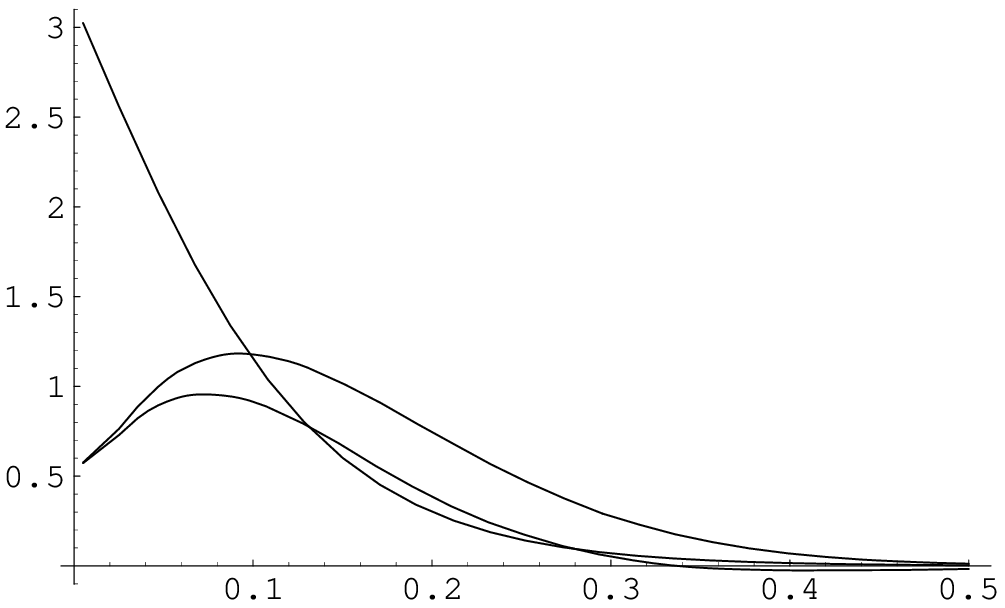, width=6cm}

{\footnotesize {\bf Figure \ref{fig3}(B).} $n_{L,R,\omega}(\rho)$ as a function of $\omega/a$ for $\tau$, $a \tau_c = 10$ and $a \rho = \pm 10$.}
\end{minipage}
\figlabel{fig3}
\vs{-.3}
\end{figure}

\section{Conclusion}

Particle creation has been discussed, as seen by non-inertial observers 
in gravitational backgrounds. The observer-dependence of the particle 
interpretation has been emphasised, and the problem of foliation 
dependence discussed. Bondi's\cite{Bondi,Bohm,Dinverno} {\it radar time} 
has been introduced, which provides an observer-dependent foliation of 
spacetime, depending only on the observers motion, 
and not an any choice of coordinates. We have argued that this 
observer-dependent foliation resolves the problem of foliation 
dependence, by uniquely connecting it to the  known observer-dependence 
of the particle concept (demonstrated by effects 
such as the Unruh\cite{Unruh,Davies} effect). The result is a 
particle interpretation which depends {\it only} 
on the motion of the observer, and on the background present, and which 
generalizes Gibbons' definition\cite{Gibb2} to arbitrary observers and 
non-stationary spacetimes. It also facilitates the definition of a number 
density operator, allowing us to calculate not just the total asymptotic 
particle creation, but also to say (with definable precision) where and when 
these particles were `created'. By incorporating the motion of 
the observer/detector, it links 
the `Bogoliubov coefficient' approach to particle creation with that provided 
by operational `detector' models, and provides a concrete answer to 
to the question ``what do particle detectors detect?'' Concrete applications 
of these definitions have 
been presented, to non-inertial observers in 1+1D Minkowski spacetime (other 
examples are presented elsewhere\cite{Dolby3}). We have shown how the thermal spectrum associated with a uniformly accelerating observer emerges as the limit of a class of `smooth turn-around' observers, none of whom have acceleration horizons.

This conference, on ``Time and Matter'', has fueled much successful 
discussion of the role of time in physics, and the conceptual tensions 
that often surround it. In this contribution I have described what I believe to 
be quite a deep connection between `time' and `matter'. That is, between 
our concept of `simultaneity', and our concepts of `particle' and `vacuum'. It 
is also hoped that some light may have been shed on the 
well-known conceptual tension between the `effectively absolute' role 
that time plays in quantum 
mechanics, and the role it plays in general relativity, where it is 
just one coordinate in a covariant theory. While the relevance and 
faintness of this 
light is for the reader to decide, the availability of radar time 
appears to suggest that there need not be any inconsistency 
between the foliation dependence of quantum mechanics, and the 
coordinate covariance of general relativity, 
provided the role of the observer is properly considered.


\begin{thebibliography}{99}
\bibitem{Schrod}
E. Schr\"{o}dinger, {\it Physica.} {\bf VI(9)}, 899 (1939).
\bibitem{Park1}
L. Parker, {\it Phys. Rev.} {\bf 183(5)}, 1057 (1969).
\bibitem{Park2}
L. Parker, {\it Phys. Rev. D} {\bf 3(2)}, 346 (1971).
\bibitem{Park3}
L. Parker, {\it Phys. Rev. Lett.} {\bf 21(8)}, 562 (1969).
\bibitem{Haw2}
S.W. Hawking, {\it Nature} {\bf 248}, 30 (1974).
\bibitem{Unruh}
W.G. Unruh, {\it Phys. Rev. D} {\bf 14}, 870 (1976).
\bibitem{Davies}
P.C.W. Davies, {\it J. Phys. A.} {\bf 8(4)}, 609 (1975).
\bibitem{GMR}
W. Greiner, B. M\"{u}ller and J. Rafelski, {\it Quantum Electrodynamics of Strong Fields} (Springer, 1985).
\bibitem{Kak}
M. Kaku, {\it Quantum Field Theory} (Oxford University Press, 1993). 
\bibitem{BD}
N.D. Birrell and P.C.W. Davies, {\it Quantum Fields in Curved Spacetime} (Cambridge University Press, 1982).
\bibitem{Full}
S.A. Fulling, {\it Aspects of Quantum Field Theory in Curved Space-Time} (Cambridge University Press, 1989).
\bibitem{Tak}
S. Takagi, {\it Prog. Theo. Phys.} Supplement No. {\bf 86} (1986).
\bibitem{Sr1}
L. Sriramkumar and T. Padmanabhan, {\it Int. J. Mod. Phys. D} {\bf 11}, 1 (2002).
\bibitem{DeWitt}
B.S. DeWitt, in {\it General Relativity,} eds. S.W. Hawking and W. Isreal (Cambridge University Press, 1979).
\bibitem{Sr2}
L. Sriramkumar, {\it Mod. Phys. Lett. A.} {\bf 14}, 1869 (1999).
\bibitem{mythesis}
C.E. Dolby, PhD Thesis. Available from 
http://www.mrao.cam.ac.uk/ \linebreak \~{}clifford/publications/abstracts/carl\_diss.html
\bibitem{Dolby}
C.E. Dolby and S.F. Gull, {\it Annals. Phys.} {\bf 293}, 189 (2001).
\bibitem{Dolby3}
C.E. Dolby and S.F. Gull, gr-qc/0207046.
\bibitem{GM1}
A.A. Grib and S.G. Mamaev, {\it Sov. J. Nuc. Phys.} {\bf 10(6)}, 722 (1970).
\bibitem{GM2}
A.A. Grib and S.G. Mamaev, {\it Sov. J. Nuc. Phys.} {\bf 14(4)}, 450 (1972).
\bibitem{MMS}
S.G. Mamaev, et. al., {\it Sov. Phys. JETP} {\bf 43(5)}, 823 (1976).
\bibitem{GMM}
A.A. Grib, et. al., {\it J. Phys. A: Math. Gen.} {\bf 13}, 2057 (1980).
\bibitem{Full2}
S.A. Fulling, {\it Gen. Rel. and Grav.} {\bf 10(10)}, 807 (1979).
\bibitem{Bondi}
H. Bondi, {\it Assumption and Myth in Physical Theory} (Cambridge University Press, 1967).
\bibitem{Bohm}
D. Bohm, {\it The Special Theory of Relativity} (W. A. Benjamin, 1965).
\bibitem{Me3}
C.E. Dolby and S.F. Gull, {\it Am. J. Phys.} {\bf 69}, 1257 (2001).
\bibitem{Gibb2}
G.W. Gibbons, {\it Comm. Math. Phys.} {\bf 44}, 245 (1975).
\bibitem{Dinverno}
R. D'Inverno, {\it Introducing Einsteins Relativity} (Oxford University Press, 1992).
\bibitem{Pauri} 
M. Pauri and M. Vallisneri, {\it Found. Phys. Lett.} {\bf 13(5)}, 401 (2000).
\bibitem{Jackiw}
R. Jackiw, Dirac Prize Lecture, Trieste, 1999. Available at hep-th/9903255.
\bibitem{Mark}
M.D. Goodsell, C.E. Dolby and S.F. Gull. In preparation.
\bibitem{Soffel} 
M. Soffel, B. M\"uller and W. Greiner, {\it Phys. Rev. D} {\bf 22}, 1935 (1980).
\end{thebibliography}
\end{document}